\documentclass[aps,prb,twocolumn,superscriptaddress,dbfloatfix]{revtex4-1}

\usepackage{bm}        
\usepackage{amsmath,amssymb,amsthm}   
\usepackage{mathtools}
\usepackage{braket}
\usepackage{tabularx}

\usepackage{hyperref}

\usepackage{grffile}
\usepackage{graphicx}
\usepackage{float}
\usepackage{color}
\usepackage{mhchem}
\newcommand*{\citen}[1]{%
	\begingroup
	\romannumeral-`\x 
	\setcitestyle{numbers}%
	\cite{#1}%
	\endgroup   
}

\begin{document}
\title{Size and Periodicity Effects on Terahertz Properties of Gammadion Metamaterials}
\author{Daniel M. Heligman}
\affiliation{Department of Physics, The Ohio State University, Columbus, OH, USA}
\author{A. M. Potts}
\affiliation{Department of Physics, The Ohio State University, Columbus, OH, USA}
\affiliation{Lake Shore Cryotronics, Westerville, OH, USA}
\affiliation{Department of Physics, University of California at Santa Barbara, Santa Barbara, CA, USA}
\author{N. Crescimanno}
\affiliation{Department of Physics, The Ohio State University, Columbus, OH, USA}
\author{M. T. Warren}
\affiliation{Department of Physics, The Ohio State University, Columbus, OH, USA}
\author{E. V. Jasper}
\affiliation{Department of Physics, The Ohio State University, Columbus, OH, USA}
\author{T. T. Mai}
\affiliation{Department of Physics, The Ohio State University, Columbus, OH, USA}
\author{R. Vald\'{e}s Aguilar}
\affiliation{Department of Physics, The Ohio State University, Columbus, OH, USA}
\date{\today}

\begin{abstract} 
We studied the effects on the terahertz transmission of gammadion metamatials by adjusting the size and periodicity of the metamaterial, while keeping the periodicity and size fixed, respectively. The terahertz transmission responses of these metamaterials were analyzed using terahertz time domain spectroscopy as well as finite difference time domain electromagnetic simulations. We have found that increasing size or periodicity results in a red shift of the resonant frequency, and that increasing size or decreasing periodicity results in the formation of a second resonant frequency. We associate these effects with the changing lengths of the microstrip lines that make up the metamaterial, as well as the changing strength of nearest-neighbor coupling between gammadions. 
\end{abstract}

\maketitle

\section{Introduction}

Electromagnetic metamaterials are structures that can exhibit non-naturally occurring properties \cite{veselago1968electrodynamics,pendry1999magnetism}. They are comprised of artificial structures that are smaller than the probing radiation wavelength; this length scale allows for the material to effectively behave as a continuous medium. One has freedom to design these structures to control the optical properties of the material, such as the effective permittivity and permeability. This leads to effects such as negative index of refraction, perfect lensing, and giant optical activity \cite{pendry2000negative,shelby2001experimental,padilla2006negative,pendry2006controlling,shalaev2007optical,soukoulis2007negative}. The last of the three, which is the property to rotate the polarization of electromagnetic radiation, can be achieved by devising a structure that is chiral, or exhibits a handedness. 

Many different chiral designs have been examined for their strength of optical activity. One of which that has generated much interest is the gammadion \cite{PhysRevB.86.035448,phua2016study,wang2009chiral,oh2015chiral,zhao2011conjugated,zarifi2013parameter,PhysRevLett.97.177401,plum2009metamaterial}. Though giant optical activity is prevalent in stacked gammadions \cite{PhysRevB.86.035448,zhao2011conjugated,zarifi2013parameter,wang2009chiral,PhysRevLett.97.177401,plum2009metamaterial}, single layer gammadions (SLGs) have shown evidence of optical activity from infrared to visible wavelengths \cite{vallius2003optical,PhysRevA.76.023811,phua2016study,kuwata2005giant}. Although transmission properties depending on the dimensions of the SLGs have been explored in the aforementioned regime, little analysis has been done on the terahertz (THz) frequency range.

Although the shape provides some interesting properties, adjusting its dimensions allows for precise tuning of these properties. Changing the size of particular features in the metamaterial can lead to shifts in their resonant frequencies\cite{o2007effects,munk2000frequency}. Increasing the distance (periodicity) of these metamaterials can reduce the absorption of these materials as well as changing how much they interact with neighboring structures. Knowing how the THz properties change with respect to adjustment of these parameters can be important for developing real world applications.  

In this work, we focus on adjusting the size the gammadions and their periodicity in a two-dimensional square array. As seen by other studies on achiral metamaterials \cite{bitzer2009lattice,sersic2009electric,singh2010strong}, varying these two dimensions affects the near-field coupling between neighbors, causing changes in the line-shape and resonant frequency. We studied the SLG metamaterial using time-domain terahertz spectroscopy (TDTS), and through finite difference time domain (FDTD) simulations using the MIT Electromagnetic Equation Propagation (MEEP) software \cite{oskooi2010meep}. Size (periodicity) was adjusted while keeping the periodicity (size) fixed to study their effects on the THz transmission (these parameters are depicted in the insets of FIG. 1 b. and d.). We find two resonant modes below 3 THz. The frequency associated with the first resonant mode red shifts with increasing size (Figure \ref{fig:size and periodicity}A), and the second resonance begins to form when the size is comparable to the periodicity (Figure \ref{fig:size and periodicity}B). Adjustment of the periodicity results in the first resonant frequency redshifting linearly at low periodicity, and asymptotically approaches what is assumed to be the fundamental resonance of a single gammadion. The increasing periodicity results in the disappearance of the second resonant frequency. 

\section*{Methods}

\begin{figure*}[t]
\includegraphics[width=.8\columnwidth]{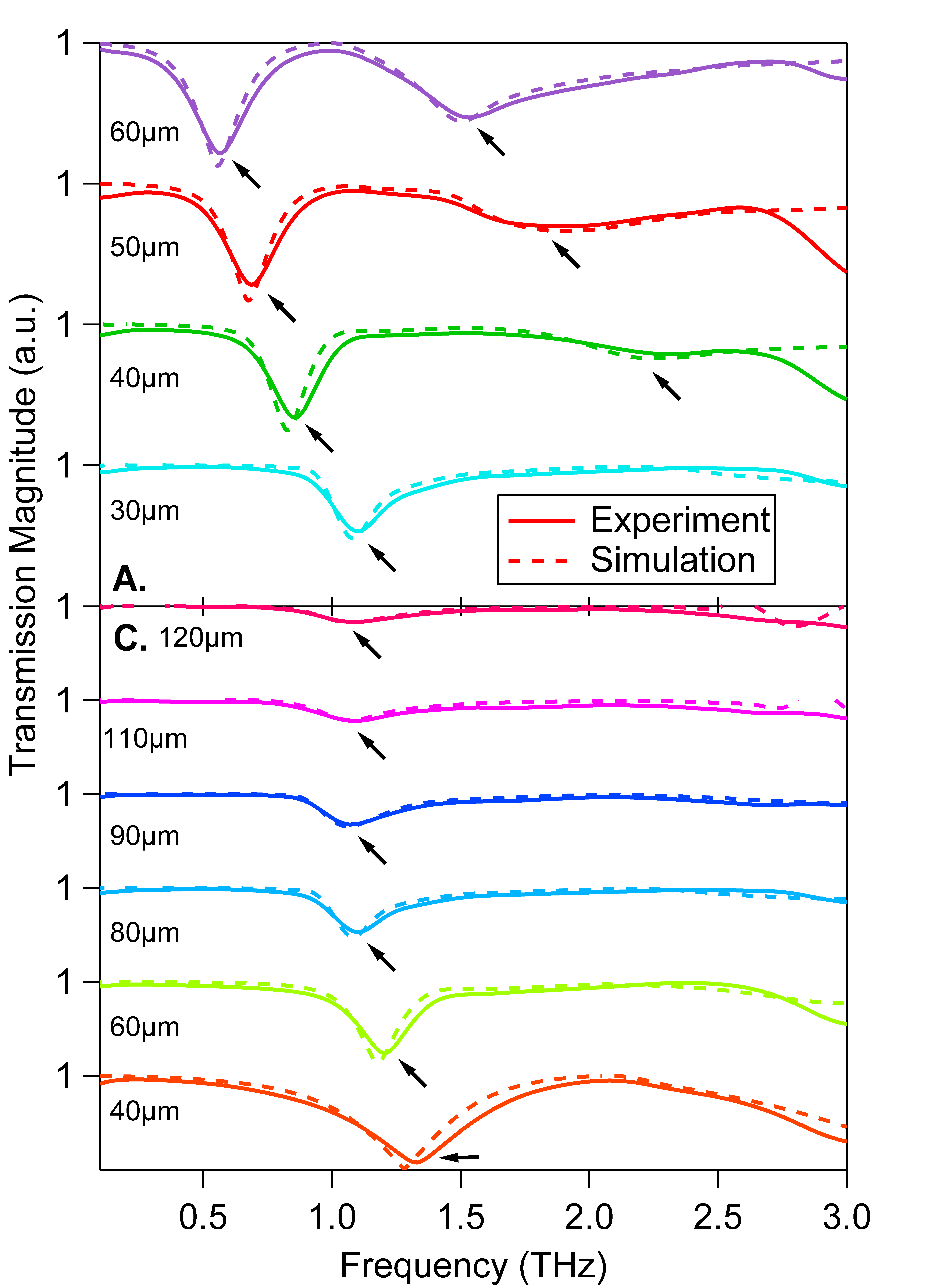}
\includegraphics[width=.8\columnwidth]{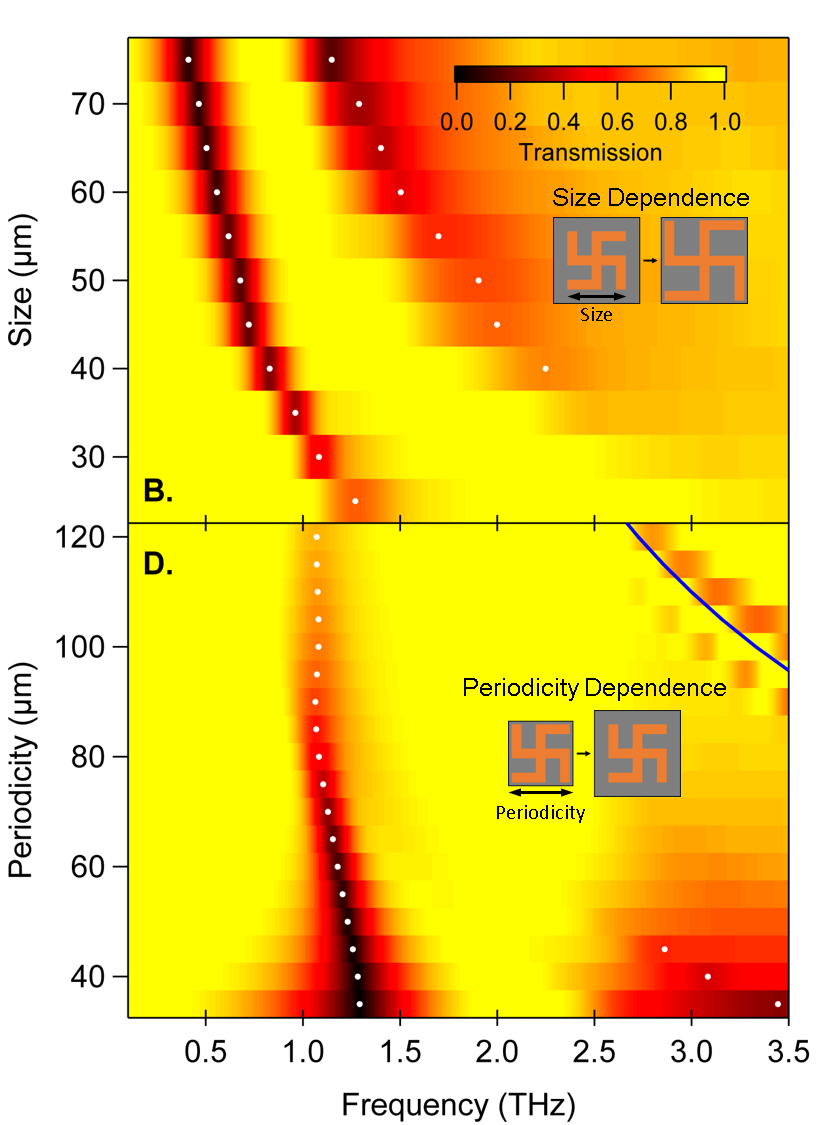}
\caption{\textbf{A.} The experimental transmission of size dependence with the periodicity fixed at 80$\mu$m. The data depicts a redshift of the resonant frequency with increasing SLG size. At large SLG size, we see the development of a second mode that becomes more defined with increasing size. This effect is viewed in more detail in simulation as depicted in \textbf{B.}, and also as dashed lines in \textbf{A.}. The data of \textbf{C.}, (periodicity dependence with fixed size of 30$\mu$m) depict a redshift of the resonant frequency with increasing periodicity accompanied by an increase in the transmission around this frequency, which is evaluated for more periodicities in simulation shown as dashed lines as well (and in panel \textbf{D.}). The blue line included in \textbf{D.} at high frequency and large periodicity, represents the diffraction mode associated with the periodicity \cite{bitzer2009lattice}. At small periodicity, the first resonant frequency tends to have a linear dependence. It is accompanied by a second resonant frequency. At high periodicity, the first resonant frequency tends to asymptote to what is assumed to be the fundamental resonance of the SLG. The aforementioned second resonance tends to die off indicating that it is a mode strictly dependent on nearest-neighbor coupling.}
\label{fig:size and periodicity}
\end{figure*}

The metamaterials used in this work were patterned using photolithography followed by a deposition of 200nm of copper on a 0.5mm thick undoped silicon substrate via physical vapor deposition. We have fabricated two sets of samples, fixed periodicity (80$\mu$m) and fixed size (30$\mu$m). The sizes of the fixed periodicity sampes ranged between 30-60$\mu$m at increments of 10$\mu$m, and for the fixed size, the periodicity ranged between 40-80$\mu$m at increments of 10$\mu$m (fig. 1a). These samples were then measured using a homebuilt\cite{PhysRevB.94.224416,warren2017terahertz,potts2017corrective} TDTS system \cite{fattinger1989terahertz,dexheimer2007terahertz}. The experimental setup utilizes a pair of IR photoconductive antennas to generate and detect terahertz pulses. The laser used to activate the antenna is an 800 nm Ti:sapphire pulsed laser which is directed to a beam splitter, one portion is sent to a receiver antenna while the other potion is sent to a retroreflector on a mechanical delay stage before reaching the emitter. Varying the position of this delay stage allows for varying the path length difference between the two beams. This permits the mapping the THz pulse in time. To ensure only linearly polarized light interacts with the sample, we placed a vertical polarizer before as well as after the sample. The metamaterials are oriented such that the E-field is parallel to one of the arms of the SLGs inner cross. Due to the 4-fold rotational symmetry of the metamaterial, any linear polarization orientation will not change the transmission spectrum\cite{ulrich1967far}. The experiment were done in a dry nitrogen environment at room temperature.

The transmission coefficient is derived by measuring the terahertz signal of a sample as well as a reference (bare silicon), Fourier-transforming them, and taking the ratio between the two. The chiral metamaterials, in general, are optically active and possess two perpendicular components: cross-polarization and co-polarization. Since our metamaterials exhibit weak optical activity, we only analyzed the co-polarized transmission. The experimental data was compared to results obtained by the numerical simulations (see figure \ref{fig:size and periodicity}). These simulations allowed for the freedom to evaluate a larger range of sizes and periodicities, which are not practical for experimental testing. In order to match the properties of the experiment, the simulated metamaterials were modeled as copper with a thickness of 200nm. The substrate was set to the index of silicon in the terahertz regime, 3.42\cite{dai2004terahertz}, assuming zero absorption. Similar to the experiment, two simulations were performed: the first was for a sample of SLGs on silicon and the second was for isolated silicon (reference). As we did experimentally, we focused on fixed periodicity at  80$\mu$m and fixed size at 30$\mu$m. The fixed periodicity set consisted of sizes ranging from 35-120$\mu$m at increments of 5$\mu$m, and the fixed size set ranged from 25-75$\mu$m with an increment of 5$\mu$m as well.

\section*{Results and Discussion}

\begin{figure}[b]
\includegraphics[width=.9\columnwidth]{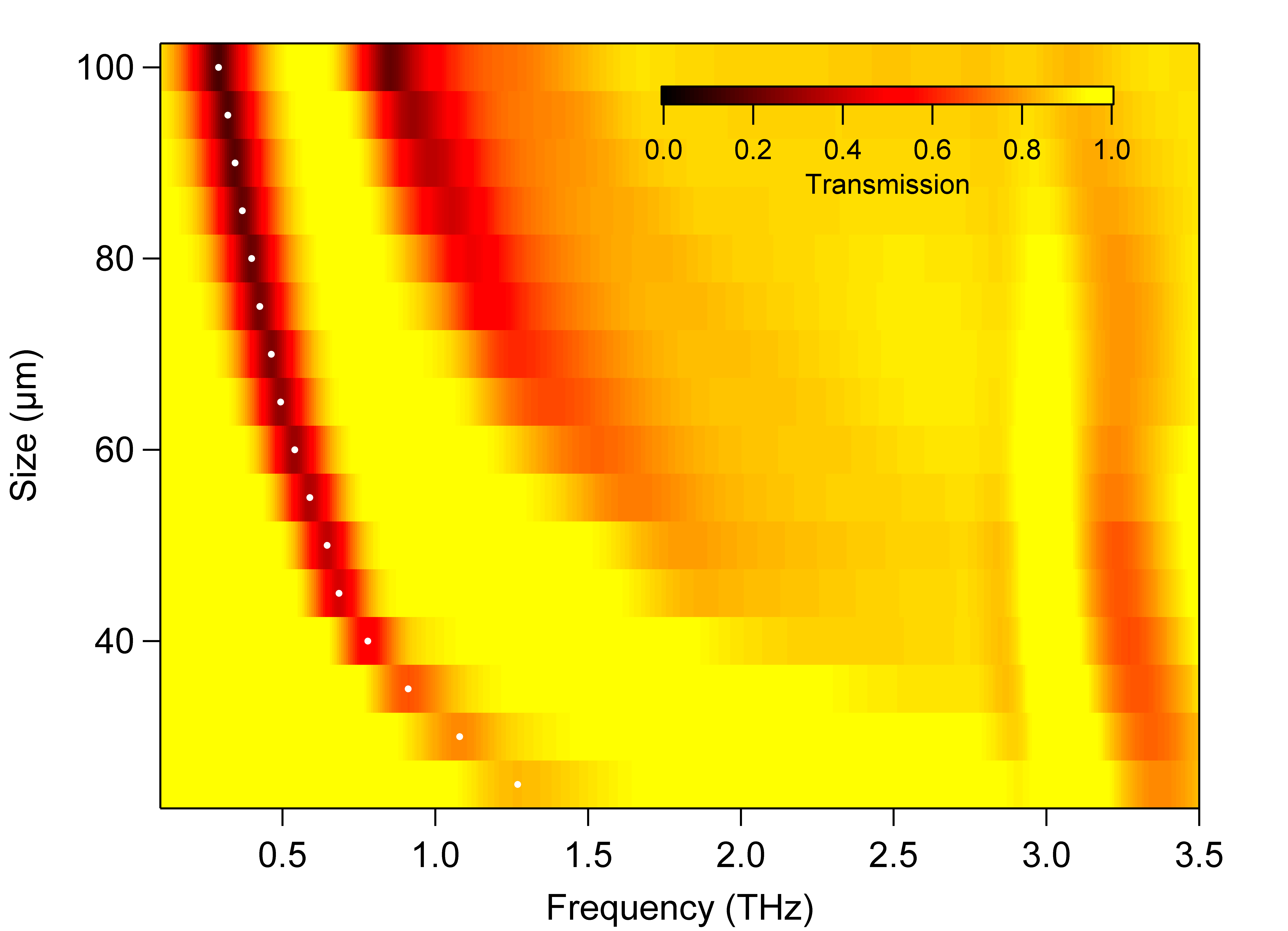}
\caption{The size dependence of SLGs with a periodicity of 105 $\mu$m. The aforementioned change in redshift rate is more pronounced for larger periodicity. The feature between 3-3.5THz is considered to be a result of diffraction from large periodicity, similar to what is observed by \citet{bitzer2009lattice}. The rapid changes in transmission above 3 THz are likely due to diffraction effects.}
\label{fig:3}
\end{figure}

\subsection{Fixed Periodicity}

The signature transmission of the SLG metamaterial involves two distinct features that we associate with resonant absorption modes as function of SLG size (Figure. \ref{fig:size and periodicity}A and B marked by arrows). As the size is increased, the resonant frequencies redshift. Within this trend, there is a significant change in the rate (slope) at which the lowest resonant frequency redshifts that occurs at a periodicity $\sim$ 40 $\mu$m. This feature is subtle in Figure \ref{fig:size and periodicity}B, but is more visible with a larger fixed periodicity as seen in Figure \ref{fig:3}, where the change in rate is seen around 45 $\mu$m. This behavior is likely due to two competing factors: the increase in size, which leads to a red shift, and the increase in nearest neighbor interactions, which, by itself, would cause a blue shift. As one increases the size of the gammadion, the length of its arms (from the center to the tip of each arm) increases, and thus it will lead to longer time-scale for the current to travel across the arm, leading to the red shift of the first resonant frequency \cite{munk2000frequency}. The electric dipole moments that arise at resonant frequency will strengthen due to the greater separation of positive and negative charges. As depicted in ref. \citen{liu2010coupling}, this could lead to an increase in the interaction energy between neighboring gammadions. This increases the restoring force between the charges which would blueshift the frequency of the resonant mode.

\begin{figure}[t]
\centering
\includegraphics[width=.9\columnwidth]{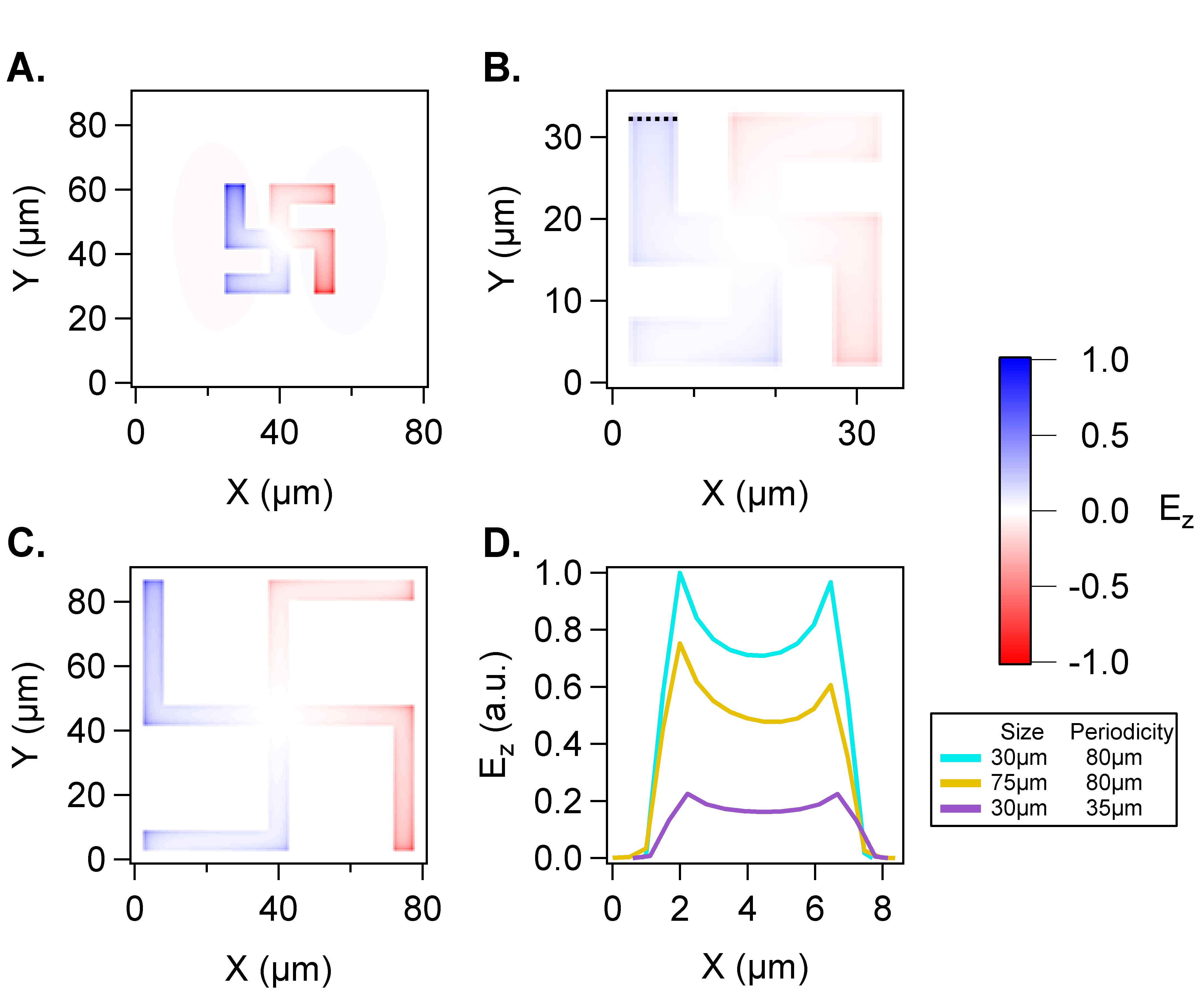}
\caption{Near fields normalized to maximum field strength in \textbf{A} for SLGs of \textbf{A} 30$\mu$m in size and 80$\mu$m in periodicity,  \textbf{B} 30$\mu$m in size and 35$\mu$m in periodicity, and \textbf{C} 75$\mu$m in size and 80$\mu$m in periodicity. The field strength for each case, \textbf{A}, \textbf{B}, and \textbf{C} were extracted along the dashed line of the SLG depicted in \textbf{B}, and the equivalent positions for the other two.  The resulting field strength is shown in \textbf{D}. The maximum field strength of \textbf{A}, \textbf{B}, and \textbf{C} are 1, $\sim$0.23, and $\sim$0.75 respectively.
}
\label{fig:2}
\end{figure}

As seen in ref. \citen{bitzer2009lattice}, an increase in the coupling strength between metamaterials, would lead to superradiant damping, which is observed as a decrease in the field strength on the metamaterial. This effect is observed in the near fields derived from FDTD simulations and confirms that coupling strength is increasing with increasing size as seen in figure \ref{fig:2}A, B and D. Here two SLGs of the same size (30 $\mu$m) but different periodicities (35 and 85 $\mu$m), show a different maximum field strength (figure \ref{fig:2}D). In particular, the SLG metamaterial with the smaller periodicity has a smaller maximum near field, indicating stronger coupling between neighbors. The combination of the redshift due to the increasing size and the blueshift due to increasing neighbor interaction energy will lead to a reduction in the rate of redshift with respect to size when neighbor interactions are large enough at a particular SLG size as seen in figure \ref{fig:size and periodicity}B. 

A second resonant frequency appeared as SLG size was increased. It does not emerge until the size of the SLGs is comparable to their periodicity, indicating that near-field interactions are connected to the formation of the second resonance. As this second resonance becomes more prominent with increasing size, we see a redshift trend similar to that of the first resonant frequency, also implying a strong size-dependence. 

\subsection{Fixed Size}

In the fixed size data set, we detected, with increasing periodicity, a redshift in the first resonant frequency (Figure \ref{fig:size and periodicity}C and D). This shift initially exhibits a linear relation to periodicity up until it is $\sim 85\mu$m. Here, the redshift abruptly stops and begins to slightly blueshift. The resonant frequency then tends to asymptote to what is assumed to be the fundamental resonant frequency of a single gammadion. Since we made no adjustments to the dimensions of the gammadion, this redshift can only be associated with the reduction of coupling strength between neighbors associated with increasing separation. As mentioned in the previous section, this coupling effect is confirmed by the reduction in near field strength. A smaller interaction energy would mean a weaker restoring force between the charges in the SLG and, therefore, a lower resonant frequency. Along with the first resonant mode there is a second one that is observed at low periodicities. As the periodicity is increased this mode tends to redshift and vanish. This behavior indicates that this second mode is not a fundamental mode of the SLG and is completely dependent on the coupling strength between neighboring structures.

\section*{Summary}
We have evaluated the effects on the THz properties of SLGs from adjusting the size (periodicity) while keeping periodicity (size) fixed. When size was increased, we observed a redshift in the first mode accompanied by the introduction of a second mode. We determined that this redshift was due to the increased timescale of the current traveling through the arms of the SLGs \cite{munk2000frequency}. With increasing size, the rate of redshifting with respect to size changed due to the introduction of increasing coupling between nearest neighbors indicating superradiant damping. From our periodicity data, we have found a redshift associated with increasing periodicity. We attributed this to the reduction of nearest neighbor coupling. With increasing periodicity we have also shown the disappearance of the second resonant frequency. We concluded from this effect that the second resonant frequency is completely associated with nearest neighbor coupling. 

\begin{acknowledgments}
This work was partially supported by OSU's Institute for Materials Research under grants IMR-FG0168 and EMR-G00030.
\end{acknowledgments}

\bibliography{Gammadionbib}

\begin{thebibliography}{33}%
\makeatletter
\providecommand \@ifxundefined [1]{%
 \@ifx{#1\undefined}
}%
\providecommand \@ifnum [1]{%
 \ifnum #1\expandafter \@firstoftwo
 \else \expandafter \@secondoftwo
 \fi
}%
\providecommand \@ifx [1]{%
 \ifx #1\expandafter \@firstoftwo
 \else \expandafter \@secondoftwo
 \fi
}%
\providecommand \natexlab [1]{#1}%
\providecommand \enquote  [1]{``#1''}%
\providecommand \bibnamefont  [1]{#1}%
\providecommand \bibfnamefont [1]{#1}%
\providecommand \citenamefont [1]{#1}%
\providecommand \href@noop [0]{\@secondoftwo}%
\providecommand \href [0]{\begingroup \@sanitize@url \@href}%
\providecommand \@href[1]{\@@startlink{#1}\@@href}%
\providecommand \@@href[1]{\endgroup#1\@@endlink}%
\providecommand \@sanitize@url [0]{\catcode `\\12\catcode `\$12\catcode
  `\&12\catcode `\#12\catcode `\^12\catcode `\_12\catcode `\%12\relax}%
\providecommand \@@startlink[1]{}%
\providecommand \@@endlink[0]{}%
\providecommand \url  [0]{\begingroup\@sanitize@url \@url }%
\providecommand \@url [1]{\endgroup\@href {#1}{\urlprefix }}%
\providecommand \urlprefix  [0]{URL }%
\providecommand \Eprint [0]{\href }%
\providecommand \doibase [0]{http://dx.doi.org/}%
\providecommand \selectlanguage [0]{\@gobble}%
\providecommand \bibinfo  [0]{\@secondoftwo}%
\providecommand \bibfield  [0]{\@secondoftwo}%
\providecommand \translation [1]{[#1]}%
\providecommand \BibitemOpen [0]{}%
\providecommand \bibitemStop [0]{}%
\providecommand \bibitemNoStop [0]{.\EOS\space}%
\providecommand \EOS [0]{\spacefactor3000\relax}%
\providecommand \BibitemShut  [1]{\csname bibitem#1\endcsname}%
\let\auto@bib@innerbib\@empty
\bibitem [{\citenamefont {Veselago}(1968)}]{veselago1968electrodynamics}%
  \BibitemOpen
  \bibfield  {author} {\bibinfo {author} {\bibfnamefont {V.~G.}\ \bibnamefont
  {Veselago}},\ }\href@noop {} {\bibfield  {journal} {\bibinfo  {journal}
  {Soviet physics uspekhi}\ }\textbf {\bibinfo {volume} {10}},\ \bibinfo
  {pages} {509} (\bibinfo {year} {1968})}\BibitemShut {NoStop}%
\bibitem [{\citenamefont {Pendry}\ \emph {et~al.}(1999)\citenamefont {Pendry},
  \citenamefont {Holden}, \citenamefont {Robbins},\ and\ \citenamefont
  {Stewart}}]{pendry1999magnetism}%
  \BibitemOpen
  \bibfield  {author} {\bibinfo {author} {\bibfnamefont {J.~B.}\ \bibnamefont
  {Pendry}}, \bibinfo {author} {\bibfnamefont {A.~J.}\ \bibnamefont {Holden}},
  \bibinfo {author} {\bibfnamefont {D.~J.}\ \bibnamefont {Robbins}}, \ and\
  \bibinfo {author} {\bibfnamefont {W.}~\bibnamefont {Stewart}},\ }\href@noop
  {} {\bibfield  {journal} {\bibinfo  {journal} {IEEE transactions on microwave
  theory and techniques}\ }\textbf {\bibinfo {volume} {47}},\ \bibinfo {pages}
  {2075} (\bibinfo {year} {1999})}\BibitemShut {NoStop}%
\bibitem [{\citenamefont {Pendry}(2000)}]{pendry2000negative}%
  \BibitemOpen
  \bibfield  {author} {\bibinfo {author} {\bibfnamefont {J.~B.}\ \bibnamefont
  {Pendry}},\ }\href@noop {} {\bibfield  {journal} {\bibinfo  {journal}
  {Physical review letters}\ }\textbf {\bibinfo {volume} {85}},\ \bibinfo
  {pages} {3966} (\bibinfo {year} {2000})}\BibitemShut {NoStop}%
\bibitem [{\citenamefont {Shelby}\ \emph {et~al.}(2001)\citenamefont {Shelby},
  \citenamefont {Smith},\ and\ \citenamefont
  {Schultz}}]{shelby2001experimental}%
  \BibitemOpen
  \bibfield  {author} {\bibinfo {author} {\bibfnamefont {R.~A.}\ \bibnamefont
  {Shelby}}, \bibinfo {author} {\bibfnamefont {D.~R.}\ \bibnamefont {Smith}}, \
  and\ \bibinfo {author} {\bibfnamefont {S.}~\bibnamefont {Schultz}},\
  }\href@noop {} {\bibfield  {journal} {\bibinfo  {journal} {science}\ }\textbf
  {\bibinfo {volume} {292}},\ \bibinfo {pages} {77} (\bibinfo {year}
  {2001})}\BibitemShut {NoStop}%
\bibitem [{\citenamefont {Padilla}\ \emph {et~al.}(2006)\citenamefont
  {Padilla}, \citenamefont {Basov},\ and\ \citenamefont
  {Smith}}]{padilla2006negative}%
  \BibitemOpen
  \bibfield  {author} {\bibinfo {author} {\bibfnamefont {W.~J.}\ \bibnamefont
  {Padilla}}, \bibinfo {author} {\bibfnamefont {D.~N.}\ \bibnamefont {Basov}},
  \ and\ \bibinfo {author} {\bibfnamefont {D.~R.}\ \bibnamefont {Smith}},\
  }\href@noop {} {\bibfield  {journal} {\bibinfo  {journal} {Materials today}\
  }\textbf {\bibinfo {volume} {9}},\ \bibinfo {pages} {28} (\bibinfo {year}
  {2006})}\BibitemShut {NoStop}%
\bibitem [{\citenamefont {Pendry}\ \emph {et~al.}(2006)\citenamefont {Pendry},
  \citenamefont {Schurig},\ and\ \citenamefont
  {Smith}}]{pendry2006controlling}%
  \BibitemOpen
  \bibfield  {author} {\bibinfo {author} {\bibfnamefont {J.~B.}\ \bibnamefont
  {Pendry}}, \bibinfo {author} {\bibfnamefont {D.}~\bibnamefont {Schurig}}, \
  and\ \bibinfo {author} {\bibfnamefont {D.~R.}\ \bibnamefont {Smith}},\
  }\href@noop {} {\bibfield  {journal} {\bibinfo  {journal} {science}\ }\textbf
  {\bibinfo {volume} {312}},\ \bibinfo {pages} {1780} (\bibinfo {year}
  {2006})}\BibitemShut {NoStop}%
\bibitem [{\citenamefont {Shalaev}(2007)}]{shalaev2007optical}%
  \BibitemOpen
  \bibfield  {author} {\bibinfo {author} {\bibfnamefont {V.~M.}\ \bibnamefont
  {Shalaev}},\ }\href@noop {} {\bibfield  {journal} {\bibinfo  {journal}
  {Nature photonics}\ }\textbf {\bibinfo {volume} {1}},\ \bibinfo {pages} {41}
  (\bibinfo {year} {2007})}\BibitemShut {NoStop}%
\bibitem [{\citenamefont {Soukoulis}\ \emph {et~al.}(2007)\citenamefont
  {Soukoulis}, \citenamefont {Linden},\ and\ \citenamefont
  {Wegener}}]{soukoulis2007negative}%
  \BibitemOpen
  \bibfield  {author} {\bibinfo {author} {\bibfnamefont {C.~M.}\ \bibnamefont
  {Soukoulis}}, \bibinfo {author} {\bibfnamefont {S.}~\bibnamefont {Linden}}, \
  and\ \bibinfo {author} {\bibfnamefont {M.}~\bibnamefont {Wegener}},\
  }\href@noop {} {\bibfield  {journal} {\bibinfo  {journal} {Science}\ }\textbf
  {\bibinfo {volume} {315}},\ \bibinfo {pages} {47} (\bibinfo {year}
  {2007})}\BibitemShut {NoStop}%
\bibitem [{\citenamefont {Zhou}\ \emph {et~al.}(2012)\citenamefont {Zhou},
  \citenamefont {Chowdhury}, \citenamefont {Zhao}, \citenamefont {Azad},
  \citenamefont {Chen}, \citenamefont {Soukoulis}, \citenamefont {Taylor},\
  and\ \citenamefont {O'Hara}}]{PhysRevB.86.035448}%
  \BibitemOpen
  \bibfield  {author} {\bibinfo {author} {\bibfnamefont {J.}~\bibnamefont
  {Zhou}}, \bibinfo {author} {\bibfnamefont {D.~R.}\ \bibnamefont {Chowdhury}},
  \bibinfo {author} {\bibfnamefont {R.}~\bibnamefont {Zhao}}, \bibinfo {author}
  {\bibfnamefont {A.~K.}\ \bibnamefont {Azad}}, \bibinfo {author}
  {\bibfnamefont {H.-T.}\ \bibnamefont {Chen}}, \bibinfo {author}
  {\bibfnamefont {C.~M.}\ \bibnamefont {Soukoulis}}, \bibinfo {author}
  {\bibfnamefont {A.~J.}\ \bibnamefont {Taylor}}, \ and\ \bibinfo {author}
  {\bibfnamefont {J.~F.}\ \bibnamefont {O'Hara}},\ }\href {\doibase
  10.1103/PhysRevB.86.035448} {\bibfield  {journal} {\bibinfo  {journal} {Phys.
  Rev. B}\ }\textbf {\bibinfo {volume} {86}},\ \bibinfo {pages} {035448}
  (\bibinfo {year} {2012})}\BibitemShut {NoStop}%
\bibitem [{\citenamefont {Phua}\ \emph {et~al.}(2016)\citenamefont {Phua},
  \citenamefont {Hor}, \citenamefont {Leong}, \citenamefont {Liu},\ and\
  \citenamefont {Khoo}}]{phua2016study}%
  \BibitemOpen
  \bibfield  {author} {\bibinfo {author} {\bibfnamefont {W.}~\bibnamefont
  {Phua}}, \bibinfo {author} {\bibfnamefont {Y.}~\bibnamefont {Hor}}, \bibinfo
  {author} {\bibfnamefont {E.~S.}\ \bibnamefont {Leong}}, \bibinfo {author}
  {\bibfnamefont {Y.}~\bibnamefont {Liu}}, \ and\ \bibinfo {author}
  {\bibfnamefont {E.}~\bibnamefont {Khoo}},\ }\href@noop {} {\bibfield
  {journal} {\bibinfo  {journal} {Plasmonics}\ }\textbf {\bibinfo {volume}
  {11}},\ \bibinfo {pages} {449} (\bibinfo {year} {2016})}\BibitemShut
  {NoStop}%
\bibitem [{\citenamefont {Wang}\ \emph {et~al.}(2009)\citenamefont {Wang},
  \citenamefont {Zhou}, \citenamefont {Koschny}, \citenamefont {Kafesaki},\
  and\ \citenamefont {Soukoulis}}]{wang2009chiral}%
  \BibitemOpen
  \bibfield  {author} {\bibinfo {author} {\bibfnamefont {B.}~\bibnamefont
  {Wang}}, \bibinfo {author} {\bibfnamefont {J.}~\bibnamefont {Zhou}}, \bibinfo
  {author} {\bibfnamefont {T.}~\bibnamefont {Koschny}}, \bibinfo {author}
  {\bibfnamefont {M.}~\bibnamefont {Kafesaki}}, \ and\ \bibinfo {author}
  {\bibfnamefont {C.~M.}\ \bibnamefont {Soukoulis}},\ }\href@noop {} {\bibfield
   {journal} {\bibinfo  {journal} {Journal of Optics A: Pure and Applied
  Optics}\ }\textbf {\bibinfo {volume} {11}},\ \bibinfo {pages} {114003}
  (\bibinfo {year} {2009})}\BibitemShut {NoStop}%
\bibitem [{\citenamefont {Oh}\ and\ \citenamefont {Hess}(2015)}]{oh2015chiral}%
  \BibitemOpen
  \bibfield  {author} {\bibinfo {author} {\bibfnamefont {S.~S.}\ \bibnamefont
  {Oh}}\ and\ \bibinfo {author} {\bibfnamefont {O.}~\bibnamefont {Hess}},\
  }\href@noop {} {\bibfield  {journal} {\bibinfo  {journal} {Nano Convergence}\
  }\textbf {\bibinfo {volume} {2}},\ \bibinfo {pages} {24} (\bibinfo {year}
  {2015})}\BibitemShut {NoStop}%
\bibitem [{\citenamefont {Zhao}\ \emph {et~al.}(2011)\citenamefont {Zhao},
  \citenamefont {Zhang}, \citenamefont {Zhou}, \citenamefont {Koschny},\ and\
  \citenamefont {Soukoulis}}]{zhao2011conjugated}%
  \BibitemOpen
  \bibfield  {author} {\bibinfo {author} {\bibfnamefont {R.}~\bibnamefont
  {Zhao}}, \bibinfo {author} {\bibfnamefont {L.}~\bibnamefont {Zhang}},
  \bibinfo {author} {\bibfnamefont {J.}~\bibnamefont {Zhou}}, \bibinfo {author}
  {\bibfnamefont {T.}~\bibnamefont {Koschny}}, \ and\ \bibinfo {author}
  {\bibfnamefont {C.}~\bibnamefont {Soukoulis}},\ }\href@noop {} {\bibfield
  {journal} {\bibinfo  {journal} {Physical Review B}\ }\textbf {\bibinfo
  {volume} {83}},\ \bibinfo {pages} {035105} (\bibinfo {year}
  {2011})}\BibitemShut {NoStop}%
\bibitem [{\citenamefont {Zarifi}\ \emph {et~al.}(2013)\citenamefont {Zarifi},
  \citenamefont {Soleimani},\ and\ \citenamefont
  {Abdolali}}]{zarifi2013parameter}%
  \BibitemOpen
  \bibfield  {author} {\bibinfo {author} {\bibfnamefont {D.}~\bibnamefont
  {Zarifi}}, \bibinfo {author} {\bibfnamefont {M.}~\bibnamefont {Soleimani}}, \
  and\ \bibinfo {author} {\bibfnamefont {A.}~\bibnamefont {Abdolali}},\
  }\href@noop {} {\bibfield  {journal} {\bibinfo  {journal} {Physical Review
  E}\ }\textbf {\bibinfo {volume} {88}},\ \bibinfo {pages} {023204} (\bibinfo
  {year} {2013})}\BibitemShut {NoStop}%
\bibitem [{\citenamefont {Rogacheva}\ \emph {et~al.}(2006)\citenamefont
  {Rogacheva}, \citenamefont {Fedotov}, \citenamefont {Schwanecke},\ and\
  \citenamefont {Zheludev}}]{PhysRevLett.97.177401}%
  \BibitemOpen
  \bibfield  {author} {\bibinfo {author} {\bibfnamefont {A.~V.}\ \bibnamefont
  {Rogacheva}}, \bibinfo {author} {\bibfnamefont {V.~A.}\ \bibnamefont
  {Fedotov}}, \bibinfo {author} {\bibfnamefont {A.~S.}\ \bibnamefont
  {Schwanecke}}, \ and\ \bibinfo {author} {\bibfnamefont {N.~I.}\ \bibnamefont
  {Zheludev}},\ }\href {\doibase 10.1103/PhysRevLett.97.177401} {\bibfield
  {journal} {\bibinfo  {journal} {Phys. Rev. Lett.}\ }\textbf {\bibinfo
  {volume} {97}},\ \bibinfo {pages} {177401} (\bibinfo {year}
  {2006})}\BibitemShut {NoStop}%
\bibitem [{\citenamefont {Plum}\ \emph {et~al.}(2009)\citenamefont {Plum},
  \citenamefont {Zhou}, \citenamefont {Dong}, \citenamefont {Fedotov},
  \citenamefont {Koschny}, \citenamefont {Soukoulis},\ and\ \citenamefont
  {Zheludev}}]{plum2009metamaterial}%
  \BibitemOpen
  \bibfield  {author} {\bibinfo {author} {\bibfnamefont {E.}~\bibnamefont
  {Plum}}, \bibinfo {author} {\bibfnamefont {J.}~\bibnamefont {Zhou}}, \bibinfo
  {author} {\bibfnamefont {J.}~\bibnamefont {Dong}}, \bibinfo {author}
  {\bibfnamefont {V.}~\bibnamefont {Fedotov}}, \bibinfo {author} {\bibfnamefont
  {T.}~\bibnamefont {Koschny}}, \bibinfo {author} {\bibfnamefont
  {C.}~\bibnamefont {Soukoulis}}, \ and\ \bibinfo {author} {\bibfnamefont
  {N.}~\bibnamefont {Zheludev}},\ }\href@noop {} {\bibfield  {journal}
  {\bibinfo  {journal} {Physical Review B}\ }\textbf {\bibinfo {volume} {79}},\
  \bibinfo {pages} {035407} (\bibinfo {year} {2009})}\BibitemShut {NoStop}%
\bibitem [{\citenamefont {Vallius}\ \emph {et~al.}(2003)\citenamefont
  {Vallius}, \citenamefont {Jefimovs}, \citenamefont {Turunen}, \citenamefont
  {Vahimaa},\ and\ \citenamefont {Svirko}}]{vallius2003optical}%
  \BibitemOpen
  \bibfield  {author} {\bibinfo {author} {\bibfnamefont {T.}~\bibnamefont
  {Vallius}}, \bibinfo {author} {\bibfnamefont {K.}~\bibnamefont {Jefimovs}},
  \bibinfo {author} {\bibfnamefont {J.}~\bibnamefont {Turunen}}, \bibinfo
  {author} {\bibfnamefont {P.}~\bibnamefont {Vahimaa}}, \ and\ \bibinfo
  {author} {\bibfnamefont {Y.}~\bibnamefont {Svirko}},\ }\href@noop {}
  {\bibfield  {journal} {\bibinfo  {journal} {Applied physics letters}\
  }\textbf {\bibinfo {volume} {83}},\ \bibinfo {pages} {234} (\bibinfo {year}
  {2003})}\BibitemShut {NoStop}%
\bibitem [{\citenamefont {Bai}\ \emph {et~al.}(2007)\citenamefont {Bai},
  \citenamefont {Svirko}, \citenamefont {Turunen},\ and\ \citenamefont
  {Vallius}}]{PhysRevA.76.023811}%
  \BibitemOpen
  \bibfield  {author} {\bibinfo {author} {\bibfnamefont {B.}~\bibnamefont
  {Bai}}, \bibinfo {author} {\bibfnamefont {Y.}~\bibnamefont {Svirko}},
  \bibinfo {author} {\bibfnamefont {J.}~\bibnamefont {Turunen}}, \ and\
  \bibinfo {author} {\bibfnamefont {T.}~\bibnamefont {Vallius}},\ }\href
  {\doibase 10.1103/PhysRevA.76.023811} {\bibfield  {journal} {\bibinfo
  {journal} {Phys. Rev. A}\ }\textbf {\bibinfo {volume} {76}},\ \bibinfo
  {pages} {023811} (\bibinfo {year} {2007})}\BibitemShut {NoStop}%
\bibitem [{\citenamefont {Kuwata-Gonokami}\ \emph {et~al.}(2005)\citenamefont
  {Kuwata-Gonokami}, \citenamefont {Saito}, \citenamefont {Ino}, \citenamefont
  {Kauranen}, \citenamefont {Jefimovs}, \citenamefont {Vallius}, \citenamefont
  {Turunen},\ and\ \citenamefont {Svirko}}]{kuwata2005giant}%
  \BibitemOpen
  \bibfield  {author} {\bibinfo {author} {\bibfnamefont {M.}~\bibnamefont
  {Kuwata-Gonokami}}, \bibinfo {author} {\bibfnamefont {N.}~\bibnamefont
  {Saito}}, \bibinfo {author} {\bibfnamefont {Y.}~\bibnamefont {Ino}}, \bibinfo
  {author} {\bibfnamefont {M.}~\bibnamefont {Kauranen}}, \bibinfo {author}
  {\bibfnamefont {K.}~\bibnamefont {Jefimovs}}, \bibinfo {author}
  {\bibfnamefont {T.}~\bibnamefont {Vallius}}, \bibinfo {author} {\bibfnamefont
  {J.}~\bibnamefont {Turunen}}, \ and\ \bibinfo {author} {\bibfnamefont
  {Y.}~\bibnamefont {Svirko}},\ }\href@noop {} {\bibfield  {journal} {\bibinfo
  {journal} {Physical review letters}\ }\textbf {\bibinfo {volume} {95}},\
  \bibinfo {pages} {227401} (\bibinfo {year} {2005})}\BibitemShut {NoStop}%
\bibitem [{\citenamefont {O'Hara}\ \emph {et~al.}(2007)\citenamefont {O'Hara},
  \citenamefont {Smirnova}, \citenamefont {Azad}, \citenamefont {Chen},\ and\
  \citenamefont {Taylor}}]{o2007effects}%
  \BibitemOpen
  \bibfield  {author} {\bibinfo {author} {\bibfnamefont {J.~F.}\ \bibnamefont
  {O'Hara}}, \bibinfo {author} {\bibfnamefont {E.}~\bibnamefont {Smirnova}},
  \bibinfo {author} {\bibfnamefont {A.~K.}\ \bibnamefont {Azad}}, \bibinfo
  {author} {\bibfnamefont {H.-T.}\ \bibnamefont {Chen}}, \ and\ \bibinfo
  {author} {\bibfnamefont {A.~J.}\ \bibnamefont {Taylor}},\ }\href@noop {}
  {\bibfield  {journal} {\bibinfo  {journal} {Active and Passive Electronic
  Components}\ }\textbf {\bibinfo {volume} {2007}} (\bibinfo {year}
  {2007})}\BibitemShut {NoStop}%
\bibitem [{\citenamefont {Munk}(2000)}]{munk2000frequency}%
  \BibitemOpen
  \bibfield  {author} {\bibinfo {author} {\bibfnamefont {B.~A.}\ \bibnamefont
  {Munk}},\ }\href@noop {} {\emph {\bibinfo {title} {Frequency selective
  surfaces: theory and design}}},\ Vol.~\bibinfo {volume} {29}\ (\bibinfo
  {publisher} {Wiley Online Library},\ \bibinfo {year} {2000})\BibitemShut
  {NoStop}%
\bibitem [{\citenamefont {Bitzer}\ \emph {et~al.}(2009)\citenamefont {Bitzer},
  \citenamefont {Wallauer}, \citenamefont {Helm}, \citenamefont {Merbold},
  \citenamefont {Feurer},\ and\ \citenamefont {Walther}}]{bitzer2009lattice}%
  \BibitemOpen
  \bibfield  {author} {\bibinfo {author} {\bibfnamefont {A.}~\bibnamefont
  {Bitzer}}, \bibinfo {author} {\bibfnamefont {J.}~\bibnamefont {Wallauer}},
  \bibinfo {author} {\bibfnamefont {H.}~\bibnamefont {Helm}}, \bibinfo {author}
  {\bibfnamefont {H.}~\bibnamefont {Merbold}}, \bibinfo {author} {\bibfnamefont
  {T.}~\bibnamefont {Feurer}}, \ and\ \bibinfo {author} {\bibfnamefont
  {M.}~\bibnamefont {Walther}},\ }\href@noop {} {\bibfield  {journal} {\bibinfo
   {journal} {Optics express}\ }\textbf {\bibinfo {volume} {17}},\ \bibinfo
  {pages} {22108} (\bibinfo {year} {2009})}\BibitemShut {NoStop}%
\bibitem [{\citenamefont {Sersic}\ \emph {et~al.}(2009)\citenamefont {Sersic},
  \citenamefont {Frimmer}, \citenamefont {Verhagen},\ and\ \citenamefont
  {Koenderink}}]{sersic2009electric}%
  \BibitemOpen
  \bibfield  {author} {\bibinfo {author} {\bibfnamefont {I.}~\bibnamefont
  {Sersic}}, \bibinfo {author} {\bibfnamefont {M.}~\bibnamefont {Frimmer}},
  \bibinfo {author} {\bibfnamefont {E.}~\bibnamefont {Verhagen}}, \ and\
  \bibinfo {author} {\bibfnamefont {A.~F.}\ \bibnamefont {Koenderink}},\
  }\href@noop {} {\bibfield  {journal} {\bibinfo  {journal} {Physical review
  letters}\ }\textbf {\bibinfo {volume} {103}},\ \bibinfo {pages} {213902}
  (\bibinfo {year} {2009})}\BibitemShut {NoStop}%
\bibitem [{\citenamefont {Singh}\ \emph {et~al.}(2010)\citenamefont {Singh},
  \citenamefont {Rockstuhl},\ and\ \citenamefont {Zhang}}]{singh2010strong}%
  \BibitemOpen
  \bibfield  {author} {\bibinfo {author} {\bibfnamefont {R.}~\bibnamefont
  {Singh}}, \bibinfo {author} {\bibfnamefont {C.}~\bibnamefont {Rockstuhl}}, \
  and\ \bibinfo {author} {\bibfnamefont {W.}~\bibnamefont {Zhang}},\
  }\href@noop {} {\bibfield  {journal} {\bibinfo  {journal} {Applied Physics
  Letters}\ }\textbf {\bibinfo {volume} {97}},\ \bibinfo {pages} {241108}
  (\bibinfo {year} {2010})}\BibitemShut {NoStop}%
\bibitem [{\citenamefont {Oskooi}\ \emph {et~al.}(2010)\citenamefont {Oskooi},
  \citenamefont {Roundy}, \citenamefont {Ibanescu}, \citenamefont {Bermel},
  \citenamefont {Joannopoulos},\ and\ \citenamefont
  {Johnson}}]{oskooi2010meep}%
  \BibitemOpen
  \bibfield  {author} {\bibinfo {author} {\bibfnamefont {A.~F.}\ \bibnamefont
  {Oskooi}}, \bibinfo {author} {\bibfnamefont {D.}~\bibnamefont {Roundy}},
  \bibinfo {author} {\bibfnamefont {M.}~\bibnamefont {Ibanescu}}, \bibinfo
  {author} {\bibfnamefont {P.}~\bibnamefont {Bermel}}, \bibinfo {author}
  {\bibfnamefont {J.~D.}\ \bibnamefont {Joannopoulos}}, \ and\ \bibinfo
  {author} {\bibfnamefont {S.~G.}\ \bibnamefont {Johnson}},\ }\href@noop {}
  {\bibfield  {journal} {\bibinfo  {journal} {Computer Physics Communications}\
  }\textbf {\bibinfo {volume} {181}},\ \bibinfo {pages} {687} (\bibinfo {year}
  {2010})}\BibitemShut {NoStop}%
\bibitem [{\citenamefont {Mai}\ \emph {et~al.}(2016)\citenamefont {Mai},
  \citenamefont {Svoboda}, \citenamefont {Warren}, \citenamefont {Jang},
  \citenamefont {Brangham}, \citenamefont {Jeong}, \citenamefont {Cheong},\
  and\ \citenamefont {Vald\'es~Aguilar}}]{PhysRevB.94.224416}%
  \BibitemOpen
  \bibfield  {author} {\bibinfo {author} {\bibfnamefont {T.~T.}\ \bibnamefont
  {Mai}}, \bibinfo {author} {\bibfnamefont {C.}~\bibnamefont {Svoboda}},
  \bibinfo {author} {\bibfnamefont {M.~T.}\ \bibnamefont {Warren}}, \bibinfo
  {author} {\bibfnamefont {T.-H.}\ \bibnamefont {Jang}}, \bibinfo {author}
  {\bibfnamefont {J.}~\bibnamefont {Brangham}}, \bibinfo {author}
  {\bibfnamefont {Y.~H.}\ \bibnamefont {Jeong}}, \bibinfo {author}
  {\bibfnamefont {S.-W.}\ \bibnamefont {Cheong}}, \ and\ \bibinfo {author}
  {\bibfnamefont {R.}~\bibnamefont {Vald\'es~Aguilar}},\ }\href {\doibase
  10.1103/PhysRevB.94.224416} {\bibfield  {journal} {\bibinfo  {journal} {Phys.
  Rev. B}\ }\textbf {\bibinfo {volume} {94}},\ \bibinfo {pages} {224416}
  (\bibinfo {year} {2016})}\BibitemShut {NoStop}%
\bibitem [{\citenamefont {Warren}\ \emph {et~al.}(2017)\citenamefont {Warren},
  \citenamefont {Pokharel}, \citenamefont {Christianson}, \citenamefont
  {Mandrus},\ and\ \citenamefont {Aguilar}}]{warren2017terahertz}%
  \BibitemOpen
  \bibfield  {author} {\bibinfo {author} {\bibfnamefont {M.~T.}\ \bibnamefont
  {Warren}}, \bibinfo {author} {\bibfnamefont {G.}~\bibnamefont {Pokharel}},
  \bibinfo {author} {\bibfnamefont {A.}~\bibnamefont {Christianson}}, \bibinfo
  {author} {\bibfnamefont {D.}~\bibnamefont {Mandrus}}, \ and\ \bibinfo
  {author} {\bibfnamefont {R.~V.}\ \bibnamefont {Aguilar}},\ }\href@noop {}
  {\bibfield  {journal} {\bibinfo  {journal} {Physical Review B}\ }\textbf
  {\bibinfo {volume} {96}},\ \bibinfo {pages} {054432} (\bibinfo {year}
  {2017})}\BibitemShut {NoStop}%
\bibitem [{\citenamefont {Potts}\ \emph {et~al.}(2017)\citenamefont {Potts},
  \citenamefont {Mai}, \citenamefont {Warren},\ and\ \citenamefont
  {Aguilar}}]{potts2017corrective}%
  \BibitemOpen
  \bibfield  {author} {\bibinfo {author} {\bibfnamefont {A.}~\bibnamefont
  {Potts}}, \bibinfo {author} {\bibfnamefont {T.}~\bibnamefont {Mai}}, \bibinfo
  {author} {\bibfnamefont {M.}~\bibnamefont {Warren}}, \ and\ \bibinfo {author}
  {\bibfnamefont {R.~V.}\ \bibnamefont {Aguilar}},\ }\href@noop {} {\bibfield
  {journal} {\bibinfo  {journal} {arXiv preprint arXiv:1712.06669}\ } (\bibinfo
  {year} {2017})}\BibitemShut {NoStop}%
\bibitem [{\citenamefont {Fattinger}\ and\ \citenamefont
  {Grischkowsky}(1989)}]{fattinger1989terahertz}%
  \BibitemOpen
  \bibfield  {author} {\bibinfo {author} {\bibfnamefont {C.}~\bibnamefont
  {Fattinger}}\ and\ \bibinfo {author} {\bibfnamefont {D.}~\bibnamefont
  {Grischkowsky}},\ }\href@noop {} {\bibfield  {journal} {\bibinfo  {journal}
  {Applied Physics Letters}\ }\textbf {\bibinfo {volume} {54}},\ \bibinfo
  {pages} {490} (\bibinfo {year} {1989})}\BibitemShut {NoStop}%
\bibitem [{\citenamefont {Dexheimer}(2007)}]{dexheimer2007terahertz}%
  \BibitemOpen
  \bibfield  {author} {\bibinfo {author} {\bibfnamefont {S.~L.}\ \bibnamefont
  {Dexheimer}},\ }\href@noop {} {\emph {\bibinfo {title} {Terahertz
  spectroscopy: principles and applications}}}\ (\bibinfo  {publisher} {CRC
  press},\ \bibinfo {year} {2007})\BibitemShut {NoStop}%
\bibitem [{\citenamefont {Ulrich}(1967)}]{ulrich1967far}%
  \BibitemOpen
  \bibfield  {author} {\bibinfo {author} {\bibfnamefont {R.}~\bibnamefont
  {Ulrich}},\ }\href@noop {} {\bibfield  {journal} {\bibinfo  {journal}
  {Infrared Physics}\ }\textbf {\bibinfo {volume} {7}},\ \bibinfo {pages} {37}
  (\bibinfo {year} {1967})}\BibitemShut {NoStop}%
\bibitem [{\citenamefont {Dai}\ \emph {et~al.}(2004)\citenamefont {Dai},
  \citenamefont {Zhang}, \citenamefont {Zhang},\ and\ \citenamefont
  {Grischkowsky}}]{dai2004terahertz}%
  \BibitemOpen
  \bibfield  {author} {\bibinfo {author} {\bibfnamefont {J.}~\bibnamefont
  {Dai}}, \bibinfo {author} {\bibfnamefont {J.}~\bibnamefont {Zhang}}, \bibinfo
  {author} {\bibfnamefont {W.}~\bibnamefont {Zhang}}, \ and\ \bibinfo {author}
  {\bibfnamefont {D.}~\bibnamefont {Grischkowsky}},\ }\href@noop {} {\bibfield
  {journal} {\bibinfo  {journal} {JOSA B}\ }\textbf {\bibinfo {volume} {21}},\
  \bibinfo {pages} {1379} (\bibinfo {year} {2004})}\BibitemShut {NoStop}%
\bibitem [{\citenamefont {Liu}\ and\ \citenamefont
  {Giessen}(2010)}]{liu2010coupling}%
  \BibitemOpen
  \bibfield  {author} {\bibinfo {author} {\bibfnamefont {N.}~\bibnamefont
  {Liu}}\ and\ \bibinfo {author} {\bibfnamefont {H.}~\bibnamefont {Giessen}},\
  }\href@noop {} {\bibfield  {journal} {\bibinfo  {journal} {Angewandte Chemie
  International Edition}\ }\textbf {\bibinfo {volume} {49}},\ \bibinfo {pages}
  {9838} (\bibinfo {year} {2010})}\BibitemShut {NoStop}%
\end{thebibliography}%
\end{document}